# Charge-Driven Liquid-Crystalline Behavior of Ligand-Functionalized Nanorods in Apolar Solvent

*Jeongmo Kim[1], Zijun Wang[1], Khalid Lahlil[1], Patrick Davidson[2*], Thierry Gacoin[1*], and Jongwook Kim[1*]*

[1]Laboratoire de Physique de la Matière Condensée, CNRS, École Polytechnique, Institut Polytechnique de Paris, 91128 Palaiseau, France

[2]Laboratoire de Physique des Solides, CNRS, Université Paris-Sud, Université Paris-Saclay, 91405 Orsay Cedex, France

*Patrick Davidson**
Full postal address : N 202, Laboratoire de Physique des Solides, CNRS, Université Paris-Sud, Université Paris-Saclay, 91405 Orsay Cedex, France
Telephone : +33 (0)1 69 15 53 93
E-mail address : patrick.davidson@u-psud.fr;

*Thierry Gacoin**
Full postal address : 04 3014C, Laboratoire de Physique de la Matière Condensée, CNRS, École Polytechnique, Institut Polytechnique de Paris, 91128 Palaiseau, France
Telephone : +33 (0)1 69 33 46 56
E-mail address : thierry.gacoin@polytechnique.edu

*Jongwook Kim**
Full postal address : 05 3066B, Laboratoire de Physique de la Matière Condensée, CNRS, École Polytechnique, Institut Polytechnique de Paris, 91128 Palaiseau, France
Telephone : +33 (0)1 69 33 46 83
E-mail address : jong-wook.kim@polytechnique.edu

**Abstract**

Concentrated colloidal suspensions of nanorods often exhibit liquid-crystalline (LC) behavior. The transition to a nematic LC phase, with long-range orientational order of the particles, is usually well captured by Onsager's theory for hard rods, at least qualitatively. The theory shows how the volume fraction at the transition decreases with increasing aspect ratio of the rods. It also explains that the long-range electrostatic repulsive interaction occurring between rods stabilized by their surface charge can significantly increase their effective diameter, resulting in a decrease of the volume fraction at the transition, as compared to sterically stabilized rods. Here, we report on a system of ligand-stabilized LaPO$_4$ nanorods, of aspect ratio ≈ 11, dispersed in apolar medium exhibiting the counter-intuitive observation that the onset of nematic self-assembly occurs at an extremely low volume fraction of ≈ 0.25%, which is lower than observed (≈ 3%) with the same particles when charge-stabilized in polar solvent. Furthermore, the nanorod volume fraction at the transition increases with increasing concentration of ligands, in a similar way as in polar media where increasing the ionic strength leads to surface-charge screening. This peculiar system was investigated by dynamic light scattering, Fourier-Transform Infra-Red spectroscopy, zetametry, electron microscopy, polarized-light microscopy, photoluminescence measurements, and X-ray scattering. Based on these experimental data, we formulate several tentative scenarios that might explain this unexpected phase behavior. However, at this stage, its full understanding remains a pending theoretical challenge. Nevertheless, this study shows that dispersing anisotropic nanoparticles in an apolar solvent may sometimes lead to spontaneous ordering events that defy our intuitive ideas about colloidal systems.

## 1  Introduction

Colloidal liquid crystals (LC) are suspensions in a solvent of small particles, with highly anisotropic shape, that exhibit, upon increasing concentration, a spontaneous transition from a disordered isotropic liquid phase to ordered yet fluid phases (*e.g.* nematic, smectic, columnar LC phases). Since the discovery of this phenomenon in early 20$^{th}$ century by Zocher using rod-like V$_2$O$_5$ nanoparticles [1], the LC behaviors of both natural systems (*e.g.* rod-like tobacco mosaic virus [2, 3], bentonite clay platelets [4]) and synthetic nanorod colloids (*e.g.* TiO$_2$ [5], γ-AlO(OH) [6, 7], α-FeOOH [8, 9], NaYF$_4$ [10], LaPO$_4$ [11], CdSe [12]) were generally shown to follow the Onsager model of the isotropic-to-nematic (I/N) transition of hard rods: the volume fraction $\Phi_{I/N}$ at which the transition takes place is inversely proportional to the aspect ratio of the rods (L/D where L is the length and D is the diameter of the rods). [13]

In polar solvents, nanorods can be stabilized by their electrical surface charges giving rise to electrostatic repulsions between the particles. This long-range repulsive interaction makes the particles appear effectively larger than their physical (steric) size. This is usually accounted for, in the frame of the Onsager model, by considering an effective diameter ($D_{eff}$) which is a key parameter controlling their LC behavior. [13] Although the effective particle aspect ratio decreases with increasing $D_{eff}$ as (L/$D_{eff}$), the effective volume fraction of rods ($\Phi_{eff} = L \cdot D_{eff}^2$) increases faster, quadratically, so that the latter effect dominates the former. Consequently, the I/N transition of charged rods usually occurs at lower volume fraction than observed for hard rods, if all other parameters are kept constant. [13] For instance, LaPO$_4$ nanorods dispersed in ethylene glycol have shown a significant decrease of $\Phi_{I/N}$ from 17% (expected with the physical rod diameter) to 3.4% (observed value) due to their high surface charge density and thus to their significantly increased $D_{eff}$. [11] In addition, variation of the ionic strength of the solvent, which can be tuned by the amount of added salts, provides an additional means of adjusting $\Phi_{I/N}$ by controlling the screening of the electrostatic repulsion between the particles in polar solvents. [11, 14, 15]

In contrast, the effect of the surface charge of nanoparticles in apolar media is usually neglected because charge dissociation requires high energy to overcome the strong Coulombic interactions that occur for small values of the dielectric constant. Stabilization of colloids in apolar media then relies primarily on the steric hindrance of molecular or polymeric ligands introduced to avoid the aggregation of particles due to short-range attractive Van der Waals interactions.[16] Steric repulsion extends typically over a few nanometers, which corresponds to the thickness of the ligand brush. [17] This thickness is usually much smaller than the electrostatic Debye screening length, $\kappa^{-1}$, of charged colloids in polar media (which is roughly related to $D_{eff}$ by $D_{eff} \approx D + 2\kappa^{-1}$) that can reach up to a few hundreds of nanometers at low ionic strength.[18] Consequently, LC phases of sterically-stabilized particles in apolar media usually display a nematic phase at relatively high volume fractions. As examples of the few systems reported so far, boehmite ($\gamma$-AlO(OH)) nanorods (L = 180 nm, L/D = 18) [7] and CdSe quantum rods (L = 40 nm, L/D = 10) [12] dispersed in cyclohexane have $\Phi_{I/N}$ of 12% and 50%, respectively. Note that preparing such highly concentrated yet well-dispersed solutions of anisotropic colloids without coping with aggregation or viscoelasticity problems is very difficult.

In this work, we describe the counter-intuitive LC behavior of a system of ligand-functionalized nanorods dispersed in an apolar medium, with $\Phi_{I/N}$ that is almost two orders of magnitude lower than predicted by the Onsager model for a sterically stabilized system. The nanorods have an average aspect ratio (L/D) of $\approx$ 11 (L = 367 nm, D = 32 nm), and are made of crystalline lanthanum phosphate doped with europium (LaPO$_4$:Eu). They are functionalized with n-Decyl Phosphonic Acid (n-DPA) and dispersed in chloroform (CHCl$_3$). We took advantage of the photoluminescence signal from the Eu$^{3+}$ dopants to accurately measure the nanorod concentration and its continuous gradient in samples held in capillaries stored vertically (i.e. under the influence of gravity). While the prediction by the Onsager model for $\Phi_{I/N}$ is $\approx$ 33% ($\Phi_I$ = 3.3D/L $\approx$ 29%; $\Phi_N$ = 4.2D/L $\approx$ 37%), we instead observed an extremely low $\Phi_{I/N}$ value of $\approx$ 0.25%, which is even much lower than observed ($\approx$ 3%) with the same, but unfunctionalized, nanorods when charge-stabilized in a polar medium.[11] Furthermore, $\Phi_{I/N}$ increases with increasing amount of excess ligands (n-DPA) dissolved in the host medium (CHCl$_3$), in a manner similar to polar media where an increase in ionic strength leads to surface-charge screening and increase of $\Phi_{I/N}$. [11] Under the influence of gravity, these dilute nanorod suspensions exhibit a concentration profile and a macroscopic phase boundary below which either a fluid nematic phase or a nematic gel appears, depending on the n-DPA concentration. Based on these findings, we discuss several scenarios that might explain the very low value of the transition volume fraction.

## 2 Materials and Methods

### 2.1 Preparation of LaPO$_4$:Eu nanorods suspensions in ethylene glycol

Bare LaPO$_4$ nanorods, with 20% Eu doping were solvothermally synthesized and dispersed in ethylene glycol by the following method, as reported in our previous work.[11] Briefly, aqueous precursor solutions of (NH$_4$)$_2$HPO$_4$ (50 mM), La(NO$_3$)$_3$·6H$_2$0 (40 mM), Eu(NO$_3$)$_3$·H$_2$0 (10 mM) were cooled down to 0 ºC and mixed, resulting in the precipitation of seed particles. A milky solution of seed particles was then heated in a sealed glass tube at 170 ºC for 3 hours. Particles were collected by centrifugation, re-dispersed in acidic water (pH 2 adjusted with HNO$_3$), and dialyzed against the same acidic water for at least two days. Purified aqueous dispersions were then transferred into ethylene glycol (EG) by distillation using a rotary pump.

### 2.2 Preparation of functionalized LaPO$_4$:Eu nanorods with n-decyl phosphonic acid

In a typical experiment, 10 mL of LaPO$_4$:Eu nanorod suspension in EG (volume fraction $\Phi$ = 0.03 %) and 10 mL of n-decyl phosphonic acid (n-DPA) dissolved in chloroform (quarterly equivalent to mole number of nanorods in EG) were poured into a vial, and the biphasic mixture was stirred slowly overnight at room temperature. The nanorods transfer from EG to chloroform due to the presence of the alkyl chains of the n-DPA ligand molecule on the particle surface. Functionalized nanorods in chloroform were then collected by centrifugation and washed with chloroform several times. The nanorods are then re-dispersed in chloroform with a controlled amount of excess n-DPA molecules. To functionalize more nanorods at once, the volume of reacting solution was increased while keeping the same nanorod volume fraction and its relative n-DPA ligand concentration. The functionalization in this case takes longer (e.g. 3 days for 100 mL of EG and chloroform) than in a typical experiment.

### 2.3 Characterizations

The hydrodynamic size and zeta-potential of nanorods were measured by dynamic light scattering (DLS) using Malvern Panalytical Zetasizer Nano ZS (**Figure 1c, Figure 2b**). A ZEN1002 dip cell was used for measuring the zeta-potential of nanorods in apolar solvent. Scanning electron microscopy (SEM) was performed using a Hitachi S4800 microscope (**Figure 1d-e**). The SEM samples were prepared by spin-coating bare or functionalized LaPO$_4$:Eu nanorod suspensions on glass substrates. For spin-coating nanorods in chloroform, the glass substrates were pre-wetted by the solvent to minimize evaporation-induced aggregation. Fourier transform infrared spectroscopy (FTIR) was performed in transmission mode using a custom-made setup (**Figure 2a**). Functionalized nanorod LC suspensions were inserted into optical cells (thickness of 6.8 μm) by capillarity and sealed with UV-cured glue. They were observed with an Olympus BX51 polarizing microscope and their birefringence textures were recorded with a CCD camera (Discovery plus DTA DX 1600E SN) (**Figure 3-4**). Functionalized nanorod LC suspensions were also filled either by capillarity or mild centrifugation into flat optical glass tubes (Vitrocom, NJ, USA) of 100 µm thickness, 2 mm width, and 10 cm length that were sealed either with a torch or with UV-cured glue. These samples were stored upright to investigate the influence of gravity on the phase behavior. After equilibrium was reached (in a few days), the LCs in glass tubes were observed using an Olympus BX51 polarizing microscope and the optical textures were recorded with a sCMEX-20 camera (Euromex, Netherlands) (**Figure 7c, 7h**). The photoluminescence signal of $Eu^{3+}$-doped $LaPO_4$ nanorods along the capillary was collected with a fiber-coupled spectrometer (SpectraPro-300i, Princeton Instrument equipped with LN/CCD-1100-OO camera) under excitation of a monochromatic laser (OXXIUS, λ = 394 nm, FWHM = 0.7 nm) (**Figure 7b, 7g**). Other samples of colloidal dispersions were prepared in cylindrical Lindemann glass capillaries of 1 mm diameter (Glas-Technik & Konstruktion, Germany) for investigation by small-angle X-ray scattering (SAXS). SAXS experiments were performed at the Swing beamline of the SOLEIL synchrotron radiation facility at Saint-Aubin, France. The X-ray wavelength was 0.1033 nm, the beamsize was 375 µm (resp. 75 µm) in the horizontal (resp. vertical) direction, and the sample-to-detection distance was 6.028 m. Scattered X-rays were detected by an Eiger 4M detector with 75 µm pixel size with 2×2 binning, and the typical exposure time was 1 s (**Figure 7d, 7i**).

## 3   Results and Discussion

### 3.1   Ligand-stabilized LaPO₄:Eu nanorods in apolar medium

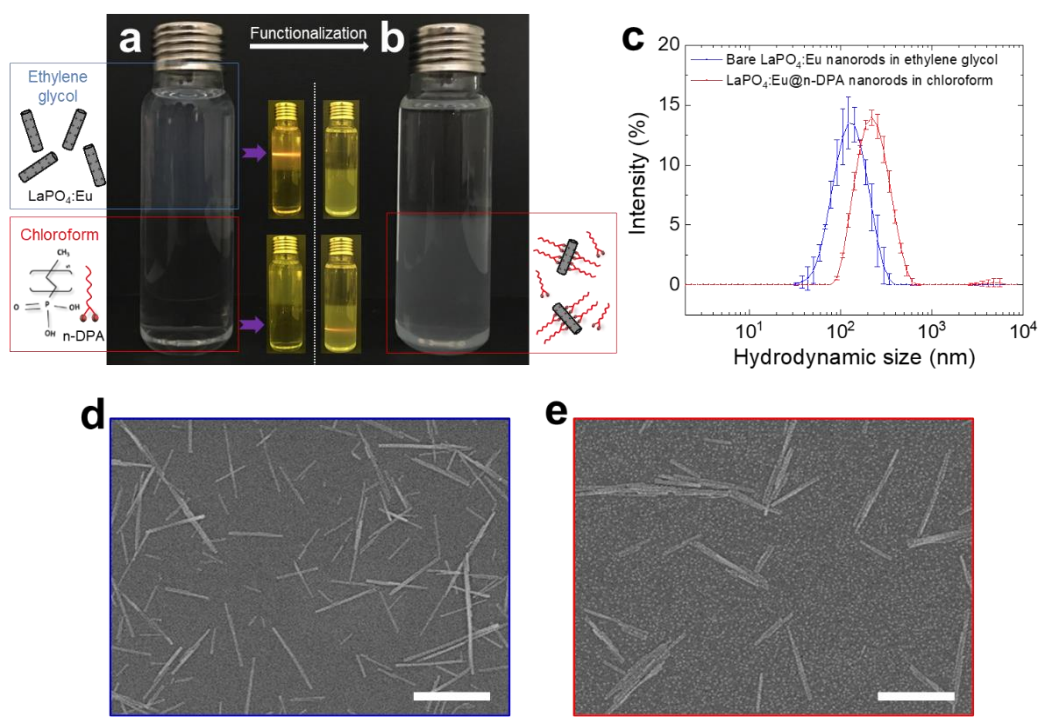

**Figure 1**. Photographs of a biphasic mixture of (top) $LaPO_4$:Eu nanorod colloidal suspension in ethylene glycol (EG) and (bottom) n-decyl phosphonic acid (n-DPA) in chloroform **(a)** before and **(b)** after the nanorods transfer into chloroform by surface-functionalization. Insets: red emission

of $Eu^{3+}$ ions doped in $LaPO_4$ nanorods under laser excitation ($\lambda_{ex}$ = 394 nm). The arrows indicate the direction of the excitation laser. A long-pass optical filter ($\lambda_{pass}$ > 510 nm) was placed in front of the camera to improve the visibility of the red emission. **(c)** Intensity plot of hydrodynamic size of the bare $LaPO_4$:Eu nanorods dispersed in EG and the $LaPO_4$:Eu@n-DPA nanorods dispersed in chloroform, measured by dynamic light scattering (DLS). **(d-e)** Scanning electron microscope (SEM) images of **(d)** the bare $LaPO_4$:Eu nanorods and **(e)** $LaPO_4$:Eu@n-DPA nanorods. (Scale bar = 500 nm).

The $LaPO_4$:Eu nanorods were solvothermally synthesized and dispersed in acidic water (pH = 2) according to the procedure described in our previous report.[11] After several days of purification by dialysis in acidic water, the nanorods were transferred into ethylene glycol (EG) where they are permanently dispersed and free of any aggregation.[11] The surface of as-prepared nanorods was then functionalized with n-decyl phosphonic acid (n-DPA) ligand molecules by mixing the dispersion of bare nanorods in EG with a n-DPA solution in chloroform. As EG and chloroform are immiscible, the functionalization reaction occurs slowly at the interface between the two phases.[19, 20] **Figure 1a** shows photographs of the initial state of the biphasic mixture where the top phase is EG and the bottom phase is chloroform. The transparent, yet slightly cloudy, aspect of the top phase indicates good dispersion of the nanorods in EG, the cloudiness arising from light scattering as the nanorod length of several hundreds of nanometers is comparable to the wavelengths of visible light. The phosphonic acid (PA) ligand moiety was chosen as a surface functionalizing group because it forms stable complexes with metal phosphates. [21] The decyl group tail was deemed ideal to provide adequate hydrophobic steric layer as well as efficient solvation, allowing for good dispersion in apolar solvents.

Upon gentle stirring of the mixture, the surface of the nanorods located at the liquid-liquid interface adsorbs the n-DPA molecules by their complexing PA groups, with the decyl chains forming a brush. Sufficiently functionalized nanorods transfer into the chloroform phase where they become more stable since they are fully functionalized. This process goes on until all the nanorods in EG have reached the liquid-liquid interface by convection or diffusion, and they are then transferred into chloroform (see **Figure S1-2** for further discussions on the functionalization process). The complete transfer of the nanorods into the organic phase was verified by the red photoluminescence of $Eu^{3+}$ dopant ions in the $LaPO_4$ structure under laser excitation at $\lambda_{ex}$ = 394 nm (**Figure 1a-b** insets). The transfer can also be assessed from the slight cloudiness of the bottom phase in **Figure 1b**. DLS measurements showed an increase of the hydrodynamic size of the nanorods, from 122 nm to 220 nm after the n-DPA functionalization and transfer (**Figure 1c),** indicating weak aggregation. Number and volume plot of hydrodynamic size comparison are shown in **Figure S3**. Note that the hydrodynamic size determined by DLS does not correspond directly to the physical size of anisotropic objects because the model used for the calculation does not take into account their highly anisotropic rod shape. Instead, this measurement allowed us to relatively assess the degree of colloidal aggregation after functionalization.

The dimensions of the nanorods are more reliably obtained from SEM measurements (**Figure 1d-e, Figure S4**). The SEM images indeed show that some weak aggregation proceeds mostly through parallel sticking of a few nanorods during the transfer process, so that the resulting thin bundles keep the rod-like shape and high aspect ratio of the individual nanorods. The actual length and diameter of the nanorods in EG and their thin aggregates in chloroform are respectively $L_{EG}$ = 155 nm and $D_{EG}$ = 19 nm before functionalization, and $L_{CHCl3}$ = 367 nm and $D_{CHCl3}$ = 32 nm after functionalization and transfer (**Figure S4**). It is worth noting that the aspect ratio of the thin bundles in chloroform remains similar to that of the individual nanorods in EG ($L_{EG}/D_{EG}$ = 8.2 and $L_{CHCl3}/D_{CHCl3}$ =10.5), so that this slight particle aggregation should not alter very much the volume fraction of the I/N transition.

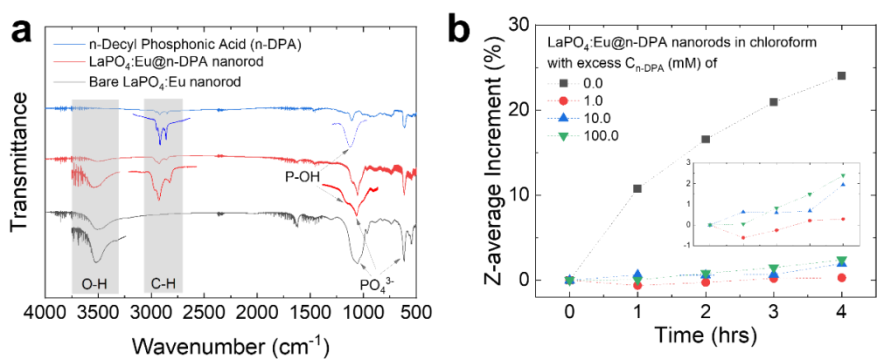

**Figure. 2 (a)** Fourier transform infrared microscopy (FTIR) spectra of free n-DPA ligand molecules, surface-functionalized $LaPO_4$:Eu@n-DPA nanorods and bare $LaPO_4$:Eu nanorods. **(b)** Evolution with time of Z-average size increments of the surface-functionalized $LaPO_4$:Eu@n-DPA nanorods dispersed in chloroform for different concentrations of excess n-DPA ($C_{n\text{-DPA}}$) of 0, 1, 10, and 100 mM. The Z-average increment (%) is calculated by dividing the difference between the Z-averages of the fresh sample and the aged sample by the Z-average of the fresh sample.

The efficiency of the complexation of the n-DPA ligands onto the nanorods was confirmed by Fourier transform infrared spectroscopy (FTIR) measurements (**Figure 2a**). The spectrum of free n-DPA molecules (blue line) shows three peaks due to the C-H stretching vibration of ordered alkyl chains at 2850 cm$^{-1}$, 2918 cm$^{-1}$, and 2957 cm$^{-1}$.[22] The functionalized nanorods were washed several times with chloroform to remove unattached n-DPA molecules. The spectrum of the functionalized nanorods (red line) shows these C-H vibrational modes at similar positions (2855 cm$^{-1}$, 2926 cm$^{-1}$, and 2956 cm$^{-1}$), confirming the presence of alkyl chains attached on the nanorod surface. The shifts of the peaks toward higher wavenumbers indicate that the alkyl chains on the particle surface are disordered.[23] The peak due to the P-OH groups (~1108 cm$^{-1}$) also demonstrates the presence of PA on the particle surface.

Subsequently, the functionalized nanorods were re-dispersed either in pure chloroform or in chloroform with controlled amounts of excess n-DPA molecules added to examine the effect of excess ligands on the colloidal properties. **Figure 2b** shows the evolution with time of the hydrodynamic Z-average increment measured by DLS. The functionalized nanorods freshly dispersed in pure chloroform ($C_{n\text{-DPA}}$ = 0.0 mM) significantly aggregate, showing around 25% of Z-average increase in a few hours. However, the functionalized nanorods with excess ligands ($C_{n\text{-DPA}}$ = 1.0 – 100.0 mM) shows a Z-average increase smaller than 2.5 % (**Figure 2b**). This result implies that the complexation of the nanorods by the n-DPA ligands is reversible and that an equilibrium between complexed and free ligand states is reached.[24, 25] Over a long period of time (typically, more than a week), the nanorods dispersed with excess n-DPA also showed an increase in Z-average size of about 50 %. But they can easily be redispersed by mild sonication, leading to the recovery of the initial hydrodynamic size, in contrast with the particles with no excess n-DPA that showed severe aggregation. This implies that the increased size of particles dispersed with no excess n-DPA ligands is most likely due to strong permanent van der Waals attractions, whereas the size increase of the particles with large excess n-DPA ligands is more likely due to transient ligand brush interactions that do not compromise much the colloidal stability. The functionalized nanorods in chloroform with excess ligands exhibit strong flow birefringence that is easily observed by shaking samples placed between crossed polarizers (**Figure S5)**. This illustrates the high stability and mobility of the individual colloids in these conditions, which is a prerequisite for LC behavior. For further studies, the nanorod volume fraction was adjusted to the desired values by dilution or slow evaporation of the solvent (chloroform).

## 3.2 Liquid-crystalline behavior

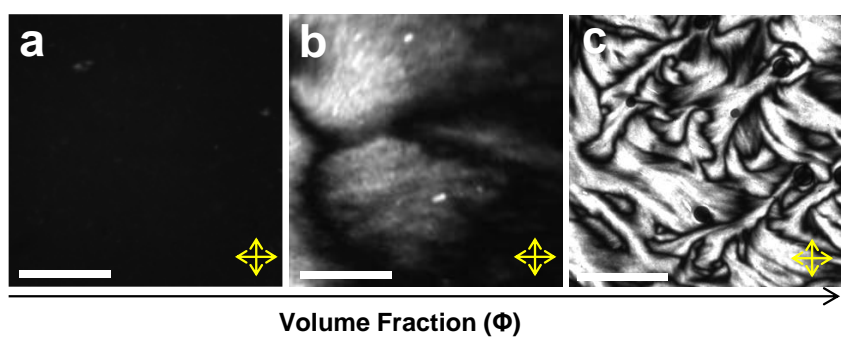

**Figure 3**. Polarized-light microscopy images showing the textures of the colloidal suspensions of functionalized LaPO$_4$:Eu@n-DPA nanorods in chloroform right after filling the optical cell: **(a)** Isotropic phase (Φ = 0.55%), showing no birefringence, **(b)** Dilute nematic phase (Φ = 0.92%) showing birefringent clouds, **(c)** Denser nematic phase showing a typical threaded texture (Φ = 2.76%). The samples were inserted into optical cells (thickness = 6.8 μm) sealed with UV-glue to prevent solvent evaporation. (The arrows show the crossed polarizers; scale bars:100 μm.)

The liquid-crystalline behavior of suspensions of n-DPA-functionalized nanorods in chloroform as a function of rod concentration was first investigated in freshly filled optical cells (thickness = 6.8 μm) lying flat on the sample stage of the polarizing microscope (here, excess n-DPA concentration in the medium is not controlled) (**Figure 3**). The suspension fills the optical cells almost instantaneously, and their birefringence texture was observed immediately after. A dilute suspension with Φ = 0.55% showed no birefringence, except for very small bright spots due to a few particle aggregates (**Figure 3a**). The nanorods thus form an isotropic phase, with random orientation, at this concentration. At slightly larger Φ, a birefringent texture appears, indicating spontaneous nematic alignment of the nanorods (**Figure 3b,** Φ = 0.92%). Rotating the stage between the crossed polarizers shows that the entire region observed under the microscope is birefringent, which means that the dispersion is homogeneous at this dilution. However, the amount of birefringence remains small, due to the low volume fraction of nanorods. At a larger volume fraction, the birefringence is stronger and a threaded texture that is a characteristic of nematic LCs is observed (**Figure 3c,** $\Phi = 2.76\%$). Interestingly, over the whole range of volume fraction explored, from a dilute and completely isotropic phase to a dense nematic phase, nematic tactoids (i.e. spindle-shaped droplets of nematic phase in coexistence with the isotropic phase)[26, 27] that are the usual sign in optical microscopy of the I/N phase transition were not detected. For this reason and because the birefringence is low, we could not precisely determine separately $\Phi_I$ and $\Phi_N$. Note nevertheless that the macroscopic I/N phase separation is readily observed in capillaries stored vertically for several days (see **Figure 7a** below).

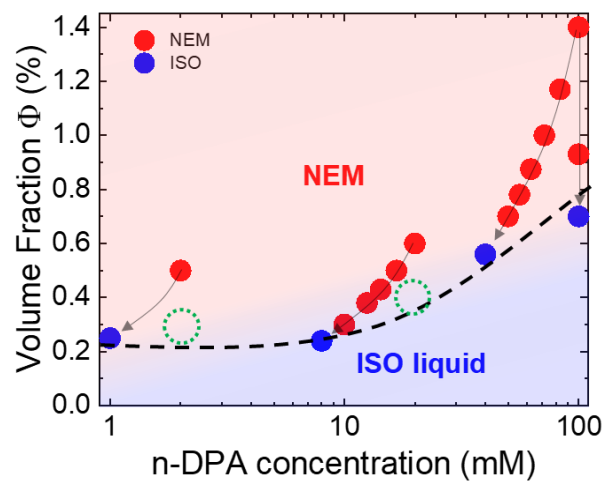

**Figure 4.** Liquid-crystal phase diagram of functionalized LaPO$_4$:Eu@n-DPA nanorods dispersed in chloroform as a function of nanorod volume fraction ($\Phi$, %) and excess n-DPA ligand concentration (mM). Red filled circles represent samples that display the nematic texture (NEM). Blue filled circles represent the first completely isotropic samples in dilution lines. The dashed line is a guide to the eye connecting the blue circles and therefore showing the isotropic part of the coexistence curve ($\Phi_I$). The green dashed empty circles indicate $\Phi_{I/N}$ data obtained from the experiment described in **Figure 7b,g**. The solid black curved arrows indicate each dilution series.

The phase diagram in **Figure 4** summarizes our observations of LC phases of the functionalized LaPO$_4$:Eu@n-DPA nanorod suspensions in chloroform as a function of nanorod volume fraction ($\Phi$, %) and excess of n-DPA ligand concentration ($C_{n\text{-DPA}}$, mM) dissolved in the medium. Due to limited amounts of colloidal material, successive dilutions of concentrated dispersions, with pure chloroform or with n-DPA solutions in chloroform at suitable concentration, were performed to obtain dilution series. In **Figure 4**, the red circles represent the experimental points where a birefringent texture appears, whereas the blue circles represent the experimental points for which the birefringent texture of the nematic phase disappears completely (lower binodal branch). The upper binodal branch that corresponds to the border between the pure nematic and the biphasic region was not precisely determined because we did not observe the I/N phase separation in the cells, as mentioned above. The polarized-light microscopy images of many samples corresponding to the data points of this diagram are shown in **Figure S6**. The LC samples represented with red symbols display a uniform birefringent texture, implying that the nematic phase has formed. The location of the isotropic to nematic transition ($\Phi_{I/N}$) and its dependence on $C_{n-DPA}$ are shown by connecting the blue circles (dashed line). Overall, qualitatively, it appears clearly that the values of $\Phi_{I/N}$ are very low, even lower than those of the same nanorods dispersed in polar media without functionalizing ligands, which is quite unexpected.[11] Moreover, the increase of $\Phi_{I/N}$ with increasing $C_{n-DPA}$ above 10 mM is similar to the effect of raising the ionic strength of LC suspensions in polar media.[11, 14]

Our observation of such a low $\Phi_{I/N}$ of the LC system in apolar solvent may be explained by several scenarios. The first type of scenario that naturally comes to mind is the influence of attractive interactions, such as depletion and van der Waals, on both the colloidal stability and the isotropic/nematic transition. On the one hand, adding excess n-DPA ligands to the dispersions could indeed provide an attractive depletion interaction to the colloids due to the formation of n-DPA micelles. However, even if all the ligands formed micelles (of typical aggregation number 100) rather than being complexed onto the nanorod surface, which is unphysical, the ligand maximum concentration of 0.1M would still lead to a depletant volume fraction of at most $10^{-2}$. Then, the depletion-induced attraction energy would remain lower than $k_BT$. [28, 29] On the other hand, even with a very high value of $10^{-18}$ J (which would already be unlikely) for the Hamaker constant of two mineral particles interacting across an organic solvent, the van der Waals attraction energy between the nanorods at 2-3 nm separation (approximate thickness of a ligand bilayer) remains rather small, at 0.25-0.5 $k_BT$.[30] Therefore, the attraction energies of both types are not strong enough to jeopardize the colloidal stability of our system. Note, by the way, that we did not observe any strong sign of colloidal aggregation, either by optical microscopy (at the macroscopic scale) or by SAXS (at the microscopic scale).

Even though these attractions are too weak to destabilize the colloidal suspensions, they still might affect the position of their isotropic/nematic transition line. Indeed, depletion interactions are known to widen the isotropic-nematic biphasic gap [31], shifting the isotropic branch of the binodal towards low concentrations. However, since the nematic branch usually also shifts towards high concentration, then the nematic phase should have a much larger volume fraction, which is not observed experimentally (see also section 3.3 below). Moreover, previous theoretical and experimental studies have shown that the very small ligand volume fraction that we use ($\leq 10^{-2}$) should not bring about depletion-induced attractions that could account for the > 10-fold shift of the transition line that is observed.[31, 32] In addition, in this scenario, increasing the ligand concentration should lower even further the onset of the phase transition [33], which contradicts the observed increase of $\Phi_{I/N}$ with increasing $C_{n-DPA}$. (Note that this argument also rules out any explanation based on a structuring effect of ligand micelles.) Other similar experimental reports of the general influence of nanoparticle attractions (not necessarily depletion-induced) on the phase diagram did not evidence any stabilization of the nematic phase to the extent that we observed. The origin of this quite unexpected phenomenon must then be of a more subtle nature.

A second type of scenario would be related to the polydispersity of the nanoparticles. Indeed, this parameter is also known to widen the isotropic-nematic biphasic gap.[34] However, the increase of the polydispersity that occurred during the n-DPA ligand functionalization (**Figure S4**) is not large enough to lower $\Phi_{I/N}$ by two orders of magnitude ($\Phi \sim 0.25\%$) compared to its expected value for sterically stabilized particles ($\Phi \sim 33\%$).

The observation of both a very low value of $\Phi_{I/N}$ and its increase with increasing $C_{n\text{-DPA}}$ could be explained in a third scenario where the nanorods experience long-range electrostatic repulsions, governed by $C_{n\text{-DPA}}$, that significantly increase their effective diameter ($D_{eff}$), even though they are dispersed in an apolar medium (chloroform). The presence of an electrical surface charge on the nanoparticles is likely the source of such long-range repulsions. Indeed, the presence of electrical charges on hydrophobic colloids (*e.g.* PMMA, TiO$_2$) in apolar media has already been clearly demonstrated in several cases.[35-38] If charge separation occured despite the apolar nature of the host media, even if the amount of charges were small, electrostatic interactions would appear and would be even longer-ranged than in polar media, due to the low dielectric constant. [38] Such long-range electrostatic repulsions would have a significant impact on the LC behavior. [15, 39]

The dielectric constant of chloroform ($\varepsilon_{r\_CHCl3} = 4.8$) is respectively 17 times and 8 times lower than those of water ($\varepsilon_{r\_water} = 80$) and ethylene glycol ($\varepsilon_{r\_EG} = 37$). The Bjerrum length $\lambda_B$ ( $\lambda_B \propto (\varepsilon_0 \varepsilon_r)^{-1}$ ) in chloroform is hence much larger ($\lambda_{B\_CHCl3} = 11.7\ nm$) than in polar solvents while the density of free ions $\rho_{ion}$ ($\rho_{ion} \propto \sqrt{e^{-\lambda_B}}$ ) is exponentially small. This results in a greatly increased Debye screening length, $\kappa^{-1}$ ( $\kappa^{-1} \propto 1/\sqrt{\lambda_B \rho_{ion}}$), in media with low dielectric constants. [38, 40, 41] In fact, it has been experimentally demonstrated that the Debye length of charged colloids can reach up to 12 μm in apolar solvents, which is about two orders of magnitude larger than in polar solvents.[18, 35, 36] Our observation of $\Phi_{I/N}$ being lower than for the same nanorods dispersed in polar solvents could then be related to very large values of $D_{eff}$ (that increases with $\kappa^{-1}$).

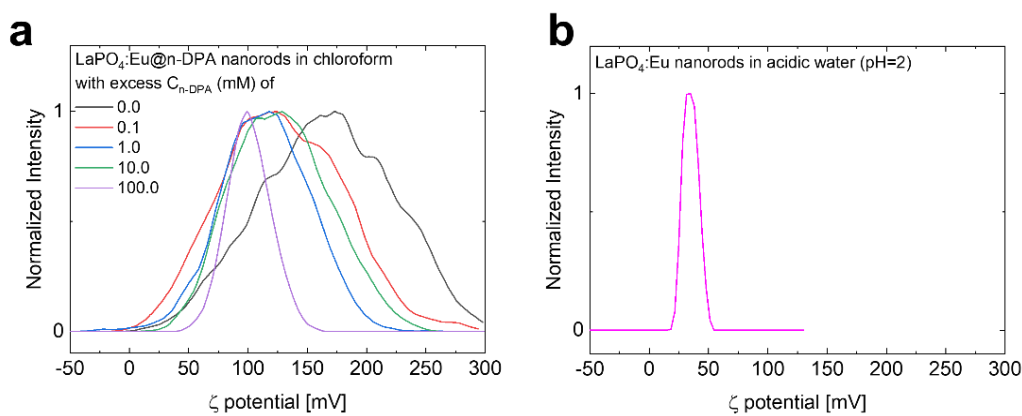

**Figure 5**. Measured zeta-potential (ζ) of **(a)** surface-functionalized LaPO$_4$:Eu@n-DPA nanorods dispersed in chloroform with different excess $C_{n\text{-DPA}}$ (mM) and **(b)** bare LaPO$_4$:Eu nanorods dispersed in acidic water (pH = 2)

Another characteristic feature of electrostatics in apolar medium is the exceptionally low double-layer capacitance $C^d$ ($C^d \propto \varepsilon_0 \varepsilon_r$) as obtained by the Debye-Hückel approximation.[42] Due to the low capacitance, even a few elementary charges can generate a sizable surface potential and strong electrostatic repulsion. [38, 40, 41] Espinosa *et al.* indeed showed that PMMA particles dispersed in solutions of nonionic sorbitan oleate surfactants in hexane are charged by no more than a few tens of elementary charges per particle while exhibiting zeta-potentials ($\zeta$) in the range of 50 - 120 mV, depending on the concentration of surfactants. [36] Inspired by this electrostatic scenario, we measured the zeta-potential of the n-DPA-functionalized LaPO$_4$:Eu nanorods in chloroform and indeed found rather large values, $\zeta \sim$ 100-150 mV, which is even higher than for the bare nanorods in aqueous medium ($\zeta \sim$ 35 mV), as shown in **Figure 5**. Moreover, $\zeta$ decreases when C$_{n\text{-DPA}}$ increases, which is consistent with the increase of $\Phi_{I/N}$ (**Figure 4** and **Figure 5a**).

From the $\zeta$ measurements above, the excess n-DPA molecules seem to act as charge control agents in the same manner as salt ions do in polar media by screening the surface charge of the nanorods. Charge control with nonionic ligands (or surfactants) in apolar media has been explained by the acid-base interaction of ligands at the particle surface and the formation of charge-carrying micelles.[36, 43, 44] **Figure 6** illustrates a 'hypothetical' chemical mechanism of the charge dissociation with n-DPA at the LaPO$_4$:Eu nanorod surface. This hypothesis involves a charge disproportionation due to acid-base interaction between the P-OH groups on the LaPO$_4$ surface and the phosphonate groups of the n-DPA ligands (**Figure 6a**). The deprotonated and negatively charged n-DPA molecules may become stable in apolar solvent by forming micelles (**Figure 6b**). The positive charges on the nanorod surface would then be separated from the anions, thus allowing for electrostatic repulsions between the rods. Since the n-DPA complexation is reversible, the dissociation probability would decrease upon increasing C$_{n\text{-DPA}}$. Therefore, adjusting C$_{n\text{-DPA}}$ should affect the screening of the protonated nanorod surface and alter the value of D$_{\text{eff}}$, in a way similar to that of the ionic strength in polar media. At this stage, the detailed investigation of the complexation dynamics and the structure of the n-DPA ligand micelles would be a considerable task that remains to be achieved in the future.

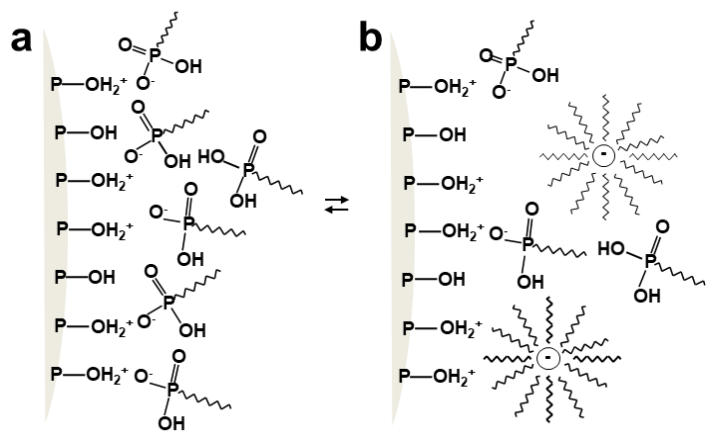

**Figure 6**. A schematic illustration of a hypothetical mechanism of charge dissociation on the LaPO$_4$:Eu nanorod surface involving **(a)** charge disproportionation of n-DPA molecules by acid-base reaction, and **(b)** micelle-like self-assembly of the deprotonated n-DPA molecules.

The explanation of the very low value of $\Phi_{I/N}$ in terms of electrostatic repulsions between particles brought about by a small surface charge appears consistent with our experimental data. However, an accurate and quantitative description of this phenomenon is presently beyond reach because the number of charges on the bare surface of nanorods and the precise value of the Debye length remain unknown. Moreover, these repulsions barely offset the van der Waals attractions taking place between the ligand-covered nanorods and giving rise to the slight particle aggregation described above (**Figure 1c**).

A last scenario is based on this slight particle aggregation. Indeed, the surface properties (such as the surface charge in the previous scenario) of the nanorods may not be uniform as their tips and lateral sides may have different surface chemistries. Then, the n-DPA ligands might not adsorb at the tips of the nanorods. This could provide a "patchy" character to the particles, which might result in their transient tip-to-tip association, leading to long nanorod chains and thus to an increase of their effective aspect ratio. Indeed, we have observed a tendency to tip-to-tip attachment of the nanorods in some samples. For example, the SEM image in **Figure S7** (of a sample in a polar solvent before functionalization) shows that multiple nanorods are tip-to-tip connected and lined-up straight. This image supports the notion of tip-to-tip attachment of the nanorods and therefore their patchy character. Such tip-to-tip attachment might occur also with ligand-functionalized particles in apolar media, which would increase their effective aspect ratio by the average number of rods per chain. This effect could help stabilizing the nematic phase at very low $\Phi$. More specific experiments would clearly be required to substantiate these conjectures.

## 3.3 Liquid-crystalline & sol-gel behaviors under gravity

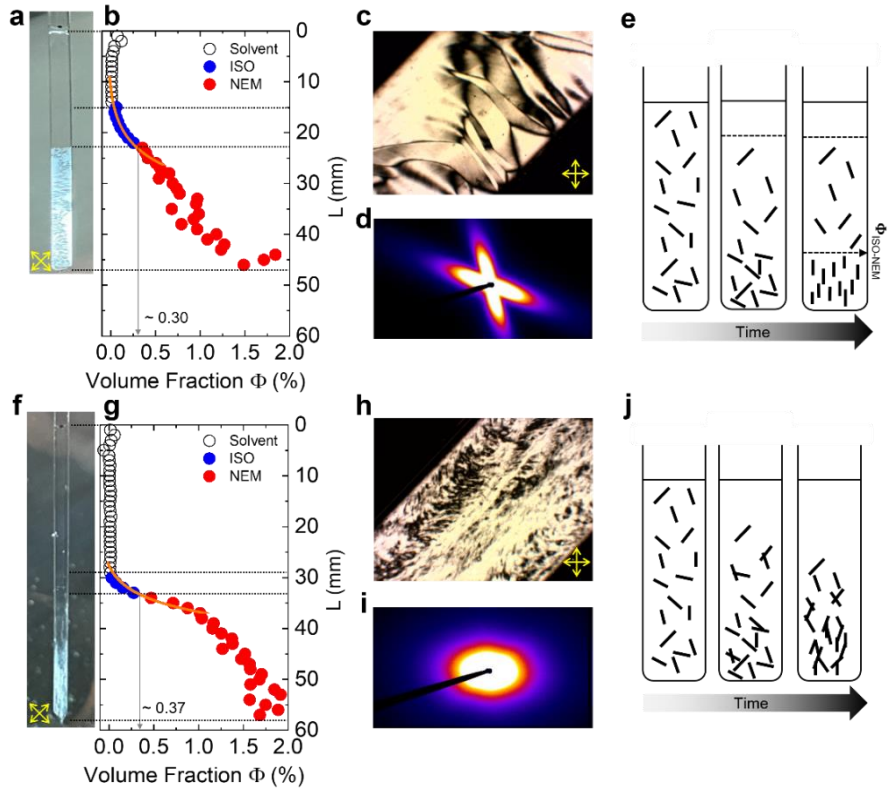

**Figure 7**. Dispersions of functionalized LaPO$_4$:Eu@n-DPA nanorods in chloroform with **(a-d)** average Φ of 0.5% and C$_{n\text{-DPA}}$ of 2.0 mM, and **(f-i)** average Φ of 0.6% and C$_{n\text{-DPA}}$ of 20.0 mM, in flat optical capillaries stored vertically. The suspensions present three different parts: pure solvent (Solvent), isotropic phase (ISO), and nematic phase (NEM). **(a,f)** Photographs of the optical capillaries placed between crossed polarizers. **(b,g)** Profiles of local volume fraction (Φ), function of L (distance measured from the liquid meniscus to the bottom of the capillary), inferred from the fluorescence intensity of Eu$^{3+}$ ions doped in the LaPO$_4$ nanorods (the emission spectra are shown in **Figure S9**). The orange line is a fit by the barometric formula (see text). **(c,h)** Polarized-light microscopy images of the nematic phases. **(d,i)** Small angle X-ray scattering (SAXS) patterns of the nematic phases. Schematic illustration of the liquid-crystalline organization of the nanorods under **(e)** barometric sedimentation and **(j)** gelation over time.

To further investigate their phase behavior, the suspensions of colloidal LaPO$_4$:Eu@n-DPA in chloroform were filled into flat glass capillaries that were flame-sealed and stored upright, in the field of gravity, for two months until equilibrium was reached (**Figure 7a,f**). When the samples are placed between crossed polarizers, a clear and sharp I/N interface was detected for samples with low C$_{n\text{-DPA}}$ (e.g. 2.0 mM, **Figure 7a**), while the transition between the isotropic and nematic phases appeared much more gradual for samples of about the same nanorod concentration but higher C$_{n\text{-DPA}}$ (e.g. 20.0 mM, **Figure 7f**). The first sample (**Figure 7c**) displayed a threaded texture typical of a fluid nematic phase. Its highly anisotropic small angle X-ray scattering (SAXS) pattern (lacking any sharp diffraction spots or lines) confirms the presence of well-aligned nematic domains (**Figure 7d**). The two scattering streaks are due to the presence of two domains with distinct nematic directors within the beam. The order parameter (**S**) calculated from each streak of the SAXS pattern is S = 0.85 ± 0.05, [45] as usually found with other dilute nematic suspensions of rod-like particles [46, 47]. In contrast, the sample with high $C_{n-DPA}$ showed a grainy birefringent texture with very small domain size (**Figure 7h**), which is typical for gels of colloidal rods subject to attractive interactions. The low anisotropy of its SAXS pattern (**Figure 7i**) also indicates the contributions of multiple small nematic domains with different orientations. High $C_{n-DPA}$ therefore seems to induce the formation of a nematic gel, as increasing the ionic strength of colloidal dispersions in polar media often does.[48] In both cases, the curve of scattered intensity versus scattering vector modulus, q, is featureless (**Figure S8**), which is common for gels but is slightly unexpected for suspensions of repulsive particles. [48] However, considering the very low concentration of the samples and the large dimensions of the scattering objects, scattering peaks would be expected around 0.01 nm$^{-1}$, which is just below the q-range covered in this experiment.

The vertical profiles of rod volume fraction (Φ) (**Figure 7b,g**) were obtained by measuring the fluorescence intensity of the nanorods doped with Eu$^{3+}$ ions (see **Figure S9** for the emission spectra). A decrease of Φ over several centimeters from the capillary bottom to the top is observed, and the I/N interface occurs within this sedimentation profile. The crudest model for the profile is an exponential dependence due to barometric equilibrium, as usually assumed for dilute suspensions of non-interacting particles. Naturally, a deviation of Φ from an exponential profile is in fact observed, which may result from particle interactions, the presence of the phase transition, gelation effects for some samples, and the coupling of the inhomogeneous concentrations of both colloids and the ions present in the suspension along the height of the capillary. [46, 49] Note that the concentration jump at the I/N transition is $\Phi_N/\Phi_I \approx 1.4$, which is actually in fair agreement with the Onsager prediction of 1.3. Besides, the gravity-induced gradient of Φ is altered by the I/N interface (**Figure 7e**) and by any interactions between colloids, as in colloidal suspensions of anisotropic particles in polar solvents. [49, 50] The dependence on $C_{n-DPA}$ of the position of the I/N interface, along the vertical axis, is noteworthy. We define the proportion of isotropic part in the sample by $L_{Iso}/L_{Iso+Nem}$, the ratio of the length of the isotropic part, $L_{Iso}$, in the capillary to the sum of the lengths of the isotropic and nematic ($L_{Nem}$) parts. This ratio is 0.23 and 0.11 for $C_{n-DPA}$ of 2.0 mM and 20.0 mM, respectively (**Figure 7b,g**). Nevertheless, very importantly, the $\Phi_{I/N}$ values measured from the capillaries stored vertically and left to phase separate under the influence of gravity match well the 2D phase diagram described previously for the freshly-prepared samples lying flat on the microscope stage (see the green dashed circles in **Figure 4**). Consequently, the sample preparation technique and history do not affect the phase diagram and the gel formation in **Figure 7f-i** appears to occur mostly after the I/N phase separation is completed.

The gravitational lengths ($l_g$) obtained by fitting the Φ-profiles around the interface with the crude barometric model: $\Phi = \Phi_0 \exp(-z/l_g)$, where $\Phi_0$ is the particle volume fraction at the bottom and z is the height [51], are 5.5 mm and 3.8 mm respectively for $C_{n-DPA}$ of 2.0 and 20.0 mM (fitted orange lines in **Figure 7b,g**). These values are larger by a factor of 1.8 and 1.3 compared to the modelled values for uncharged particles of 3 mm ($l_g = k_B T/V_p \Delta \rho g$, with k$_B$ the Boltzmann constant, T the absolute temperature, V$_p$ the particle volume, Δρ the density contrast between particle and medium, and g the gravitational acceleration) [51]. As mentioned above, this discrepancy could be related to some long-range repulsion between charged particles that is neglected here. Then, the decrease of $l_g$ with increasing $C_{n-DPA}$ could be explained by the charge screening effect as reported previously.[49, 52, 53] In addition, the deviation of the sedimentation profile of the sample with high $C_{n-DPA}$ from the exponential law in the birefringent region (**Figure 7g**) seems to be due to the gelation of the nanorods, which suppresses their mobility and prevents them from reaching thermodynamic equilibrium.

# 4 Conclusion

In summary, we have optimized the surface functionalization process of LaPO$_4$:Eu nanorods to transfer them from polar solvent (ethylene glycol) to apolar solvent (chloroform) by complexation with n-decane phosphonic acid (n-DPA), an amphiphilic molecular ligand. We have found that dilute dispersions display flow-birefringence whereas more concentrated ones show permanent nematic orientational order of the nanorods at very low volume fraction (< 1%), which is even lower than observed with dispersions of the same nanorods in EG, a polar solvent [11]. This finding was quite unexpected in colloid science because dispersions of anisotropic nanoparticles usually undergo the I/N transition at lower volume fractions when charge-stabilized in polar solvents than when sterically-stabilized in apolar solvents [7, 11, 12]. We have also found that the concentration of excess n-DPA ligands affects the zeta-potential of the nanorods and therefore their phase behavior, in a way quite similar to that of the ionic strength in aqueous dispersions, even though the system is formulated here in an apolar solvent.

Various scenarios were examined to account for the whole set of experimental observations, and the most likely one, presently, is that electrostatic repulsions between particles might explain the occurrence of a nematic phase at unexpectedly low volume fraction. Indeed, in spite of the low dielectric constant of chloroform, the LaPO$_4$:Eu@n-DPA nanorods could bear a small electric surface charge that could be due to a disproportionation reaction between the nanorod surface groups and the ligands [35-38]. Then, the nanorods would experience electrostatic repulsions decaying over a range even longer than in polar solvents [38, 40, 41]. Alternatively or perhaps even combined with the electrostatic scenario, the nanorods could also be patchy, as suggested by some TEM images showing tip-to-tip particle aggregation. At this point, in order to reach a more quantitative description of our observations, additional experimental studies are required. These include the complete characterization of the properties of the nanorods, such as their zeta potential, surface electric charge, and ligand coverage. Moreover, elucidating in detail the physical mechanism at play in this system remains a pending theoretical challenge.

This study illustrates how, contrary to popular belief, the transfer of colloidal anisotropic particles to apolar solvents, did not destabilize at all the nematic phase with respect to the isotropic one. On the contrary, orientational order appeared at a volume fraction even lower than in polar solvents. This can be exploited in practical situations where nanoparticle alignment improves material properties. Such situations arise when manufacturing ordered nanocomposites with various plastic polymer media from particles dispersed in volatile organic solvent or when preparing compact self-assembled nanostructures by evaporation or spray-assisted alignment [54-56]. In particular, aligned LaPO$_4$:Eu nanorods exhibiting strongly polarized photoluminescence can be used as anisotropic light sources [57-60].

## CRediT authorship contribution statement

**Jeongmo Kim**: Conceptualization, Investigation, Data curation, Formal analysis, Writing – Original Draft, Writing – Review & Editing. **Zijun Wang**: Investigation. **Khalid Lahlil**: Investigation, Resources. **Patrick Davidson**: Investigation, Formal analysis, Supervision, Writing – Review & Editing. **Thierry Gacoin**: Conceptualization, Supervision, Resources, Writing – Review & Editing, Funding acquisition. **Jongwook Kim**: Conceptualization, Supervision, Resources, Writing – Review & Editing, Funding acquisition.

## Data availability

The main data supporting the results are available within the article and its Supplementary Information files. Extra data will be made available from the corresponding authors on request.

## Note

This article is dedicated to the Festschrift honoring Philip (Fyl) Pincus, in recognition of his scientific achievements and profound influence.

## Declaration of competing interest

The authors declare no competing financial interest.

## Acknowledgements

This research was supported by the French national research agency (ANR-19-CE09-0033), Fondation pour la Recherche Médicale (DCM20181039556), and Institut Polytechnique de Paris (2020MATU0376). Authors would like to thank Ivan Dozov for helpful discussions and Thomas Bizien for assistance in using beamline SWING; they also acknowledge SOLEIL for provision of synchrotron radiation facilities (under the approved proposal # 20190467) and KIST Europe for support on electron microscopy.

Supporting Informations for

# Charge-Driven Liquid-Crystalline Behavior of Ligand-Functionalized Nanorods in Apolar Solvent

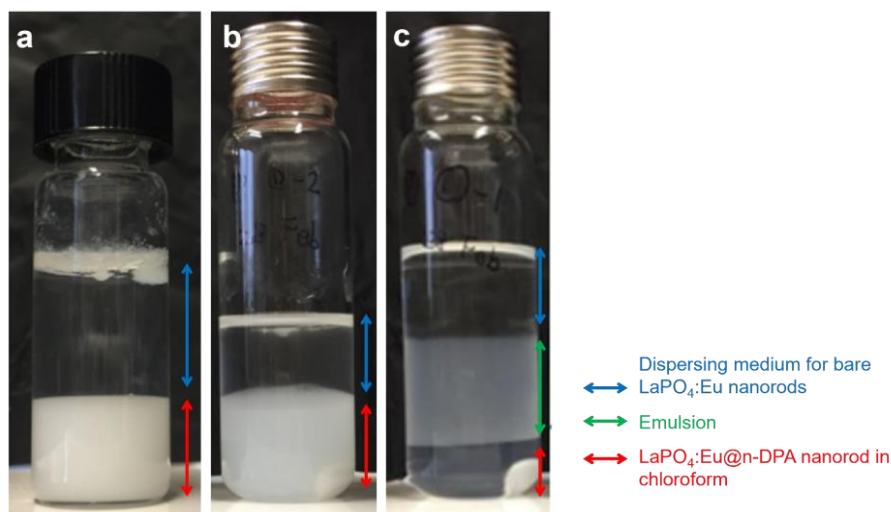

**Figure S1**. Photographs of biphasic mixtures of (top) dispersing medium of LaPO$_4$:Eu nanorods and (bottom) chloroform after the transfer of nanorods into chloroform by surface-functionalization. The dispersing medium in which LaPO$_4$:Eu nanorods were initially dispersed is (**a**) pH = 2 acidic water solution, (**b**) mixture of ethylene glycol and pH = 2 water (1:1 v/v) and (**c**) ethylene glycol.

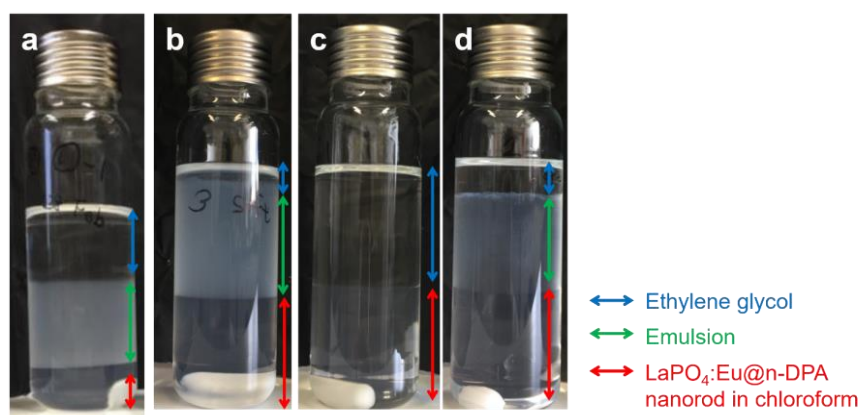

**Figure S2.** Photographs of biphasic mixtures of (top) ethylene glycol and (bottom) chloroform after the transfer of nanorods into chloroform by surface-functionalization. The volume fraction of LaPO$_4$:Eu nanorods initially dispersed in ethylene glycol (EG) and the relative amount of n-DPA molecules dissolved in chloroform are controlled. (**a**) 0.1 vol% of nanorods in EG/ 0.25 eq of n-DPA in chloroform, (**b**) 0.01 vol% of nanorods in EG/ 0.5 eq of n-DPA in chloroform, (**c**) 0.01 vol% of nanorods in EG/0.25 eq of n-DPA in chloroform, (**d**) 0.01 vol% of nanorods in EG/ 0.125 eq of n-DPA in chloroform. For each reaction, 1eq for n-DPA indicates the total mole number of nanorods that were dispersed in ethylene glycol.

There were two crucial factors to achieve efficient surface-functionalization with minimum degree of aggregation. The first step was to choose the solvent. Among the solvents evaluated as a dispersion medium for bare LaPO$_4$:Eu nanorods, it is demonstrated here that ethylene glycol results in the least degree of aggregation. (**Figure S1c**). The functionalized nanorods dispersed in other media exhibited severe irreversible aggregation, resulting in a milky look of the dispersions of surface-functionalized nanorods in chloroform, which is typical of unstable colloidal dispersions (**Figure S1a-S1b**). However, even when ethylene glycol was used as a dispersion medium (**Figure S1c**), it was found that a large number of functionalized nanorods were caught in the emulsion (mixture of ethylene glycol and chloroform) and these nanorods could not easily diffuse to the pure chloroform. The formation of emulsion was minimized by adjusting the number of LaPO$_4$:Eu nanorods and the proportion of n-DPA ligand molecules. At the optimal condition where 0.01 vol% of LaPO$_4$:Eu nanorods (1 eq in Mol) and 0.25 eq (in Mol) of n-DPA are used, the emulsion was not observed. (**Figure S2c**).

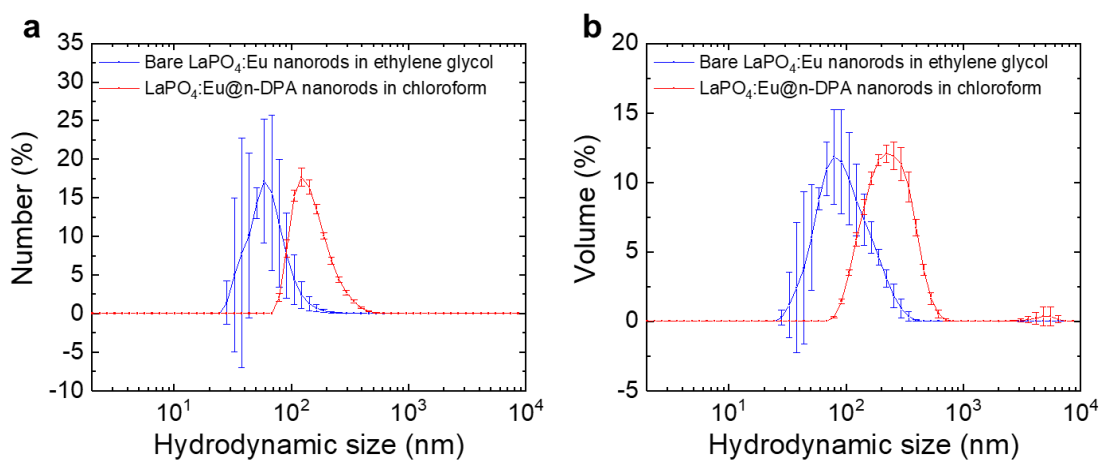

**Figure S3**. (**a**) Number and (**b**) volume plot of hydrodynamic size of the bare LaPO$_4$:Eu nanorods dispersed in EG and the LaPO$_4$:Eu@n-DPA nanorods dispersed in chloroform, measured by dynamic light scattering (DLS).

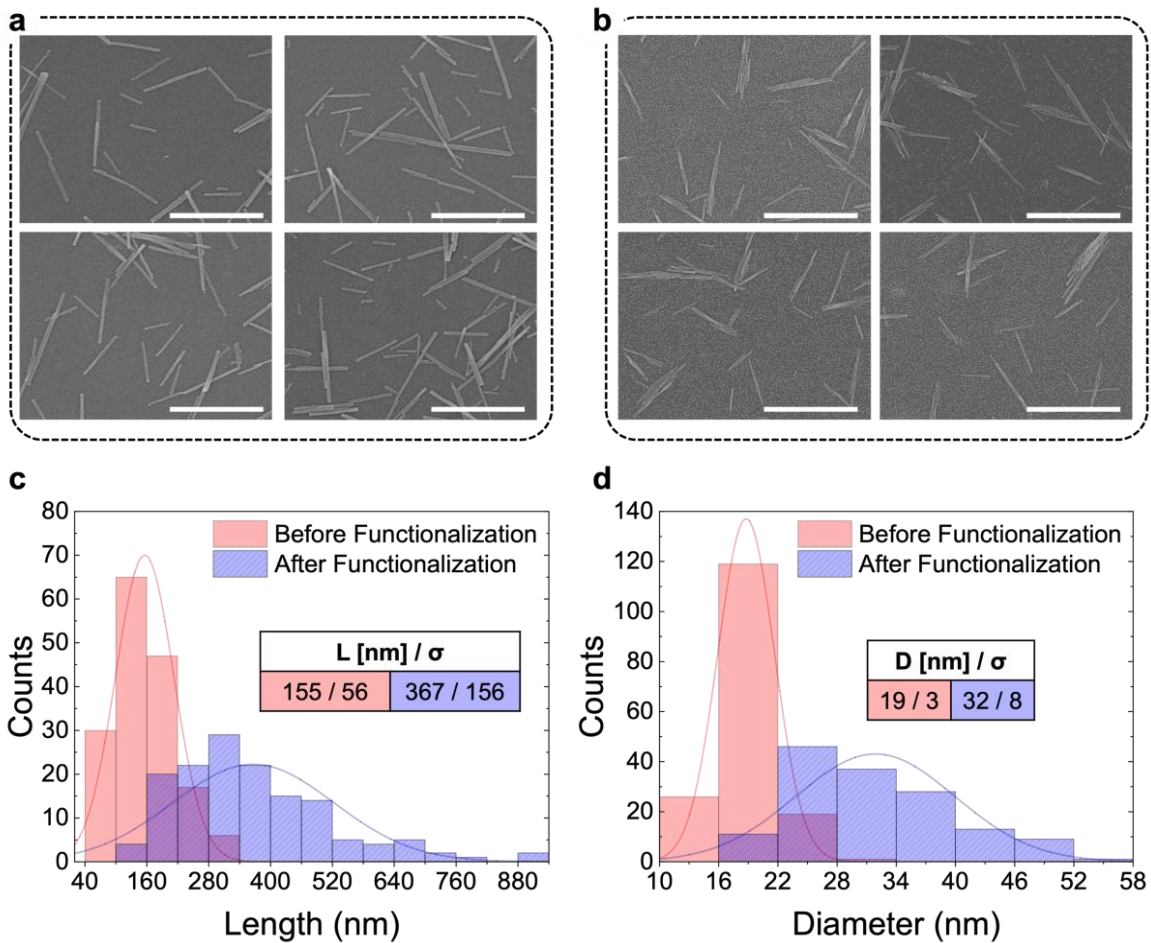

**Figure S4. (a-b)** Scanning electron microscope (SEM) images of **(a)** the bare LaPO$_4$:Eu nanorods (Scale bar = 500 nm) and **(b)** LaPO$_4$:Eu@n-DPA nanorods. (Scale bar = 1.0 μm). Measured **(c)** length and **(d)** diameter distributions of LaPO$_4$:Eu nanorods from SEM images before (in ethylene glycol) and after (in chloroform) surface-functionalization. The inset tables show the average length (L) and diameter (D) of LaPO$_4$:Eu nanorods with their standard deviations (σ).

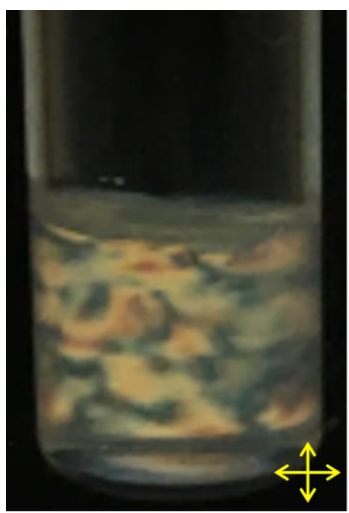

**Figure S5**. Flow birefringence of functionalized LaPO$_4$:Eu@n-DPA nanorods in chloroform (Φ = 0.55%) observed between crossed polarizers. The arrows indicate the crossed polarizers.

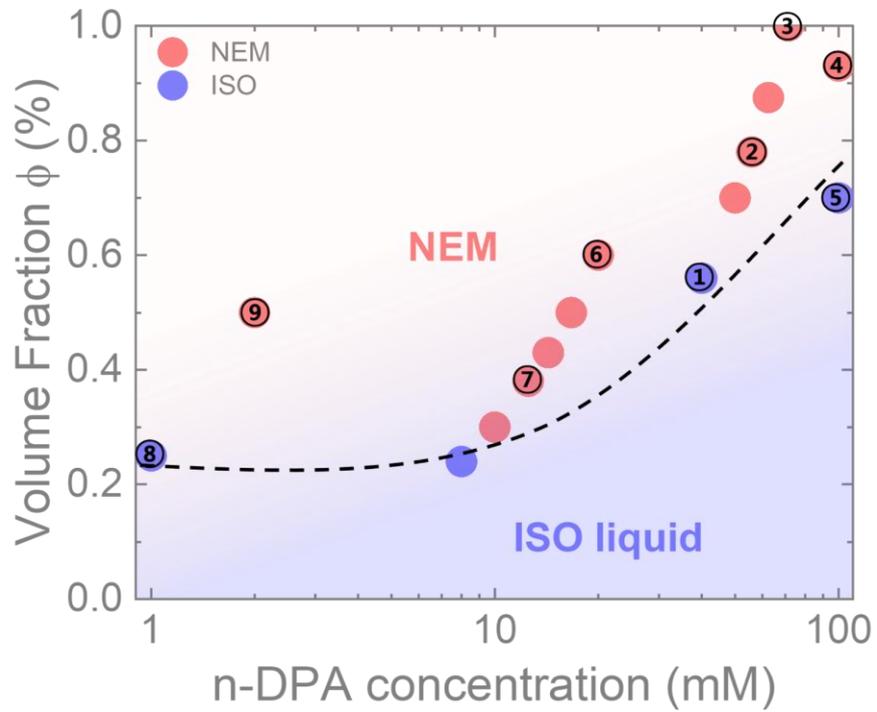

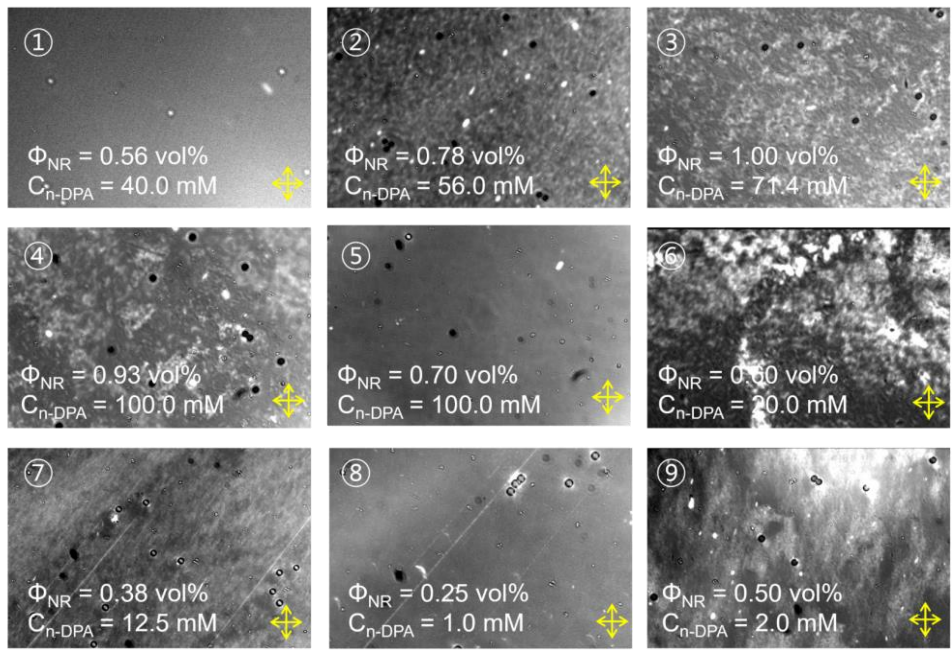

**Figure S6.** Optical texture observed between crossed polarizer of surface-functionalized LaPO$_4$:Eu@n-DPA nanorod suspension in chloroform as a function of volume fraction ($\Phi_{NR}$, vol%) and excess n-DPA ligand concentration ($C_{n\text{-}DPA}$, mM).

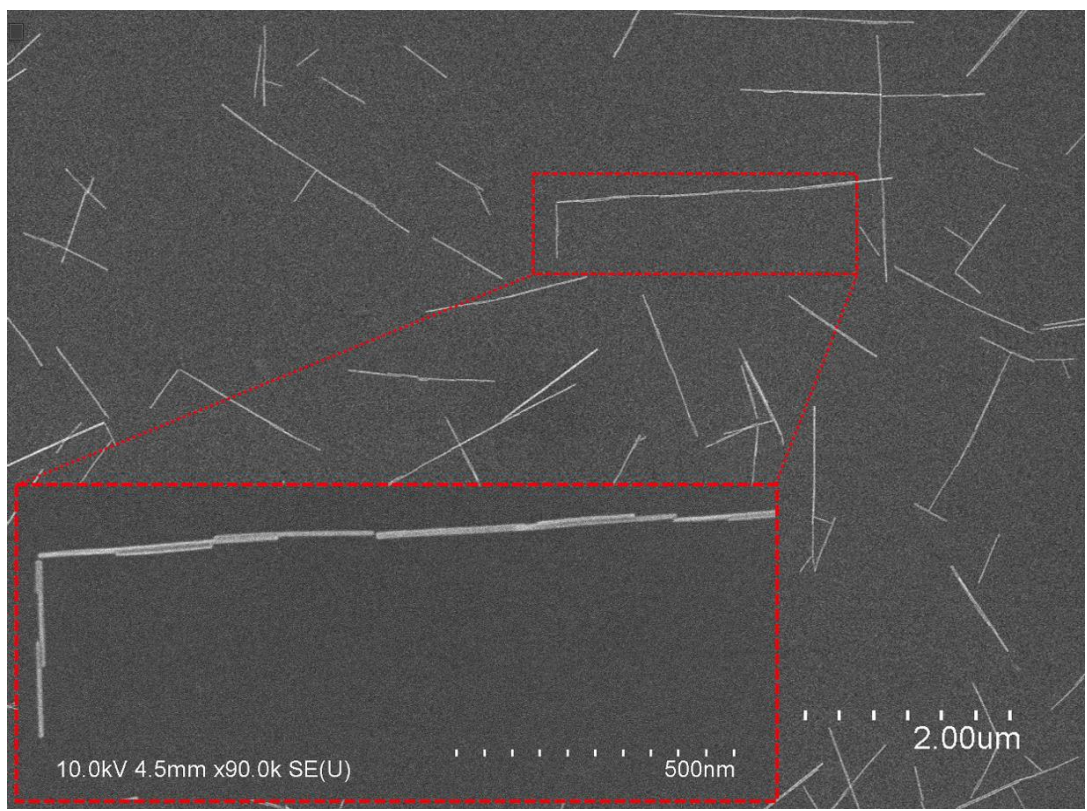

**Figure S7**. A SEM (scanning electron microscopy) image of tip-to-tip attachment of LaPO$_4$:Eu nanorods initially dispersed in polar solvent.

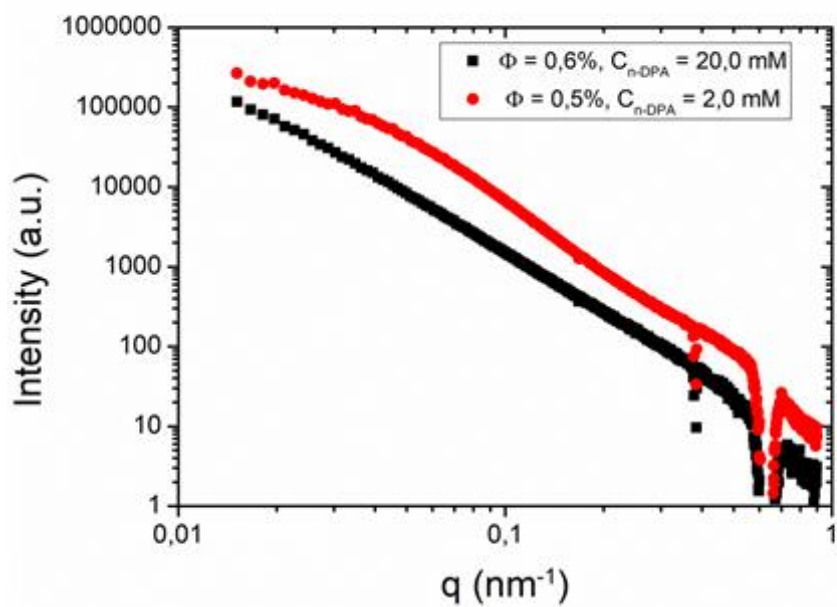

**Figure S8.** Curves of scattered intensity (in arbitrary units) versus scattering vector modulus, q, for two suspensions of surface-functionalized LaPO$_4$:Eu@n-DPA nanorods dispersed in chloroform with $\Phi = 0.5\%$, $C_{n-DPA} = 2.0$ mM (red symbols) and $\Phi = 0.6\%$, $C_{n-DPA} = 20.0$ mM (black symbols). The accidents around 0.4 and 0.6-0.7 nm$^{-1}$ are only artefacts due to the overlap of the X-ray detector tiles.

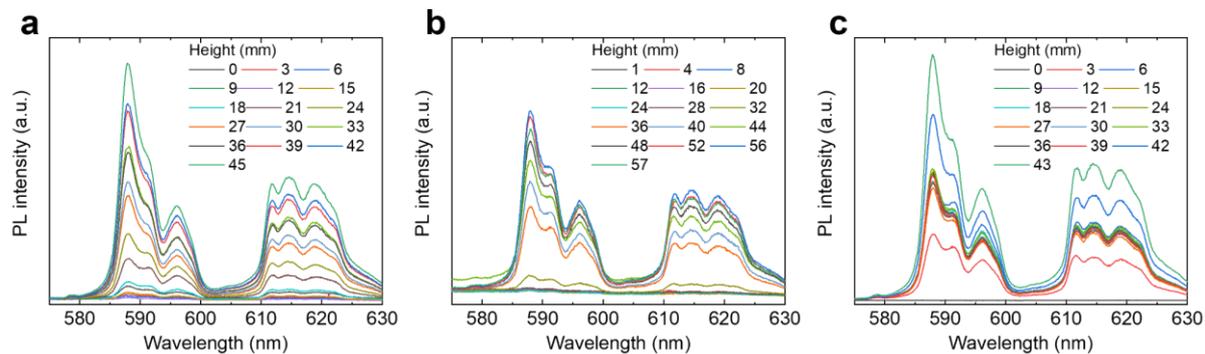

**Figure S9.** Emission spectra of $Eu^{3+}$ doped in $LaPO_4$ nanorods measured along the height of the capillary for (**a**) average $\Phi_{NR}$ = 0.5 vol% / $C_{n\text{-DPA}}$ = 2.0 mM, (**b**) average $\Phi_{NR}$ = 0.6 vol% / $C_{n\text{-DPA}}$ = 20.0 mM, (**c**) average $\Phi_{NR}$ = 1.4 vol% / $C_{n\text{-DPA}}$ = 100.0 mM.